\documentclass[]{article}

\usepackage{mathrsfs}
\usepackage{pifont}
\usepackage{amssymb}
\usepackage{graphicx}   % for figures, if needed
\begin{document}
\def\d{\hbox{d}}
\def\e{\hbox{e}}

\title{\bf{The gravitational field of a global monopole}}

\author {{Xin Shi and Xin-zhou Li} \\ \normalsize{\textit{Shanghai United Center for Astrophysics, Shanghai}}\\
\normalsize{\textit{Normal University,} 100\textit{ Guilin Road},
\textit{Shanghai} 200234, \textit{China} }}

\date{ }
\maketitle
\begin{abstract}
We present an exact solution to the non-linear equation which
describes a global monopole in the flat space. We re-examine the
metric and the geodesics outside the global monopole. We will see
that a global monopole produces a repulsive gravitational field
outside the core in addition to a solid angular deficit. The lensing
property of the global monopole and the global monopole-antimonopole
annihilation mechanism are studied.

 (This paper published in Class. Quantum Garv. \textbf{8 }(1991)
 761-767)
\end{abstract}

\section{Introduction}
\noindent The idea that magnetic monopoles ought to exist has proved
to be remarkably durable. A persuasive argument was first put
forward by Dirac [1]. Many years later, another very good argument
emerged. The 't Hooft-Polyakov monopole solution is a synthesis of
the Dirac monopole in the non-Abelian gauge theories [2],[3]. The
monopole solutions of the realistic grand unified models based on
the gauge groups $SU(5)$ and $SO(10)$ have been considered in [4]
and [5]. An extensive list of the Kaluza-Klein monopole literatures
was made in $[6]-[11]$. Surprising and qualitative phenomena arise
when one considers the quantum mechanics of electrically charged
fermions interacting with magnetic monopoles $[12]-[16]$.

 Recently, Barriola and Vilenkin [17] have shown an approximate solution of the Einstein
  equations for the metric outside a global monopole, resulting from a global symmetry breaking.
  Such a monopole has Goldstone fields with energy density decreasing with the distance only as
 $r^{-2}$, so that the total energy is linearly divergent at large distance. Neglecting the mass
 term, the monopole metric describes a space with a deficit solid angle. The area of a sphere of radius
 $r$ is not $ 4\pi r^{2}$, but $4\pi(1-8\pi G\eta^{2})r^{2}$.
  Requiring that the mass density in such a monopole should not greatly exceed the critical
  density implies there is at most one global monopole
  in the local group of galaxies. Equally stringent bounds are also
  derived which do not depend on cosmological assumptions, using the
  large tidal gravitational forces associated with the global
  monopole [18].

  In this paper, we present an exact solution to the non-linear
   equation which describes the global monopole in the flat space.
   We re-examine the metric outside a global monopole. We will see a
   repulsive gravitational field outside the core in addition to a
   solid angular deficit. We examine the geodesics and find that the
   deflected angle is small. For ultra-relativistic particles, the
   deflected angle also depends upon the impact parameter. If the
   impact parameter is of a galaxy scale and $\delta $ is of the grand
   unification scale, the term with impact parameter can be
   neglected. However, if the term is of the scale of a mini
   soliton star [19], this term is important. The lensing property
   of the global monopole, as well as other classical effects, are
   studied. We show that the repulsive gravitation force between
   global monopole and antimonopole ($M\overline{M}$) does not change the
   $M\overline{M}$
   annihilation proposed by Barriola and Vilenkin [17].

\section{An exact solution of the global monopole in flat space}
\noindent The simplest model that gives rise to the global monopole
is described by the Lagrangian
    \begin{equation}
 \mathscr{L}=\frac{1}{2}\partial_{\mu}\phi^{a}\partial^{\mu}\phi^{a}-\frac{1}{4}\lambda(\phi^{a}\phi^{a}-\eta^{2})^{2}
\end{equation}
 where $\phi^{a}$ is a triplet of scalar fields, $a=1,2,3$. The model has a global $O(3)$ symmetry which
 is spontaneously broken
 to $U(1)$. The field configuration describing a monopole is
 \begin{equation}
 \phi^{a}=\eta f(r)\frac{x^{a}}{r}
  \end{equation}The field equations for $\phi^{a}$ in the flat metric reduce to a single equation
  for $f(r)$
\begin{equation}\label{3a}
  f'' + \frac{2}{r}f'-\frac{2}{r^{2}}f-\frac{f(f^{2}-1)}{\delta^{2}}=0
  \end{equation} where
  $\delta=(\eta\sqrt{\lambda})^{-1}$
   is the core radius of the monopole. The function $f(r)$ grows linearly when
 $r<\delta $ and exponentially approaches unity as soon as $r\geq\delta.$ Barriola and Vilenkin [17] took
 $f=1$
 outside the core
 which is an approximation to the exact solution.

 On the other hand, Lan and Wang [20] have discussed an exact soliton solution of the ice-like structure in the form of
 a series of hyperbolic functions. In similar ways, we will give an exact solution of the global monopole. Furthermore, we also discuss the convergence of these series solutions.

 Setting $x=r/{\delta}$
 in equation (\ref{3a}), we obtain
   \begin{equation}\label{4a}
   x^{2}f'' +2xf'-x^{2}f^{3}+(x^{2}-2)f=0
   \end{equation}
  which satisfies the boundary conditions
   \begin{eqnarray}
   &&f(0)=0\label{5a}\ ,\\
   &&f(\infty)=1\label{5b}.
  \end{eqnarray}
This problem can be solved explicitly by the method of hyperbolic
functions, in the form
   \begin{equation}\label{6a}
   f(x)=\sum_{n=0}^{\infty}c_{n}\tanh^{2n+1}\frac{x}{\sqrt{2}}
   \end{equation}
    which satisfies the boundary condition at
  $x\rightarrow\infty$
   \begin{equation}\label{7a}
   \sum_{n=0} ^{\infty}c_{n} =1.
   \end{equation}
   By using the formula
   \begin{equation}
   x=\sqrt{2}\tanh\frac{x}{\sqrt{2}}\sum_{n=0}^{\infty}\tanh^{2n}\frac{x}{\sqrt{2}}(2n+1)^{-1},
   \end{equation}
   we have
    \begin{equation}\label{9a}
x^{2}=2\tanh^{2}\frac{x}{\sqrt{2}}\sum_{n=0}^{\infty}\left(\frac{1}{n+1}\sum_{l=0}^{\infty}
\frac{1}{2l+1}\right)\tanh^{2n}\frac{x}{\sqrt{2}}\quad .
 \end{equation}
For the determination of $c_{n}$, we first substitute equations
(\ref{6a}), (\ref{7a}) and (\ref{9a}) into equation (\ref{4a}). We
show that the recursion formula of the coefficients $c_{n}$ can be
expressed as follows:
\begin{eqnarray}\label{10a}
n(2n+3)c_{n}&=&\sum_{l+k=n-1}\left(\frac{2(2k+1)}{(2l+1)(2l+3)}
+4k(k+1)d_{l}-(2k+1)kd_{l-1}\right. \nonumber
\\ & &-(k+1)(2k+1)d_{l-1}\bigg)
c_{k}+\sum_{i+j+k+l=n-2}\!\!d_{l}c_{i}c_{j}c_{k},
\end{eqnarray}
where
\begin{equation}\label{lla}
d_{l}=\left\{\begin{array}{ll}
0 & \textrm{if \quad} l= -1\\
\frac{1}{l+1} \sum\limits_{i=0}^{l} \frac{1}{2i+1} \ & \textrm{if}
\quad l=0,1,2,3,\ldots .
\end{array}\right.
\end{equation}
Equations (\ref{7a}) and (\ref{10a}) lead to a simple method to
determine $c_{n}$.

 By using equation (\ref{7a}), we have $|f(x)|<1$ for $x\geq 0$.
 The series of hyperbolic functions (\ref{6a}) converges uniformly in
 the region $x\geq 0$. Thus we believe that the function series
(\ref{6a}) shows an exact solution of equation (\ref{4a}).

Next, we consider the approximate solution of $Nth$ order
\begin{equation}\label{12a}
 f^{(N)}=\sum_{n=0}
^{N}c_{n}^{(N)}\tanh^{2n+1}\frac{x}{\sqrt{2}}
\end{equation}
which satisfies the infinite boundary condition,
\begin{equation}\label{13a}
\sum_{n=0}^{N}c_{n}^{(N)}=1.
\end{equation}
The approximate solution of zero order is
\begin{equation}\label{28a}
f^{(0)}=\tanh\frac{x}{\sqrt{2}}.
\end{equation}
 The approximate solution of 10th order is
{\setlength\arraycolsep{2pt}
\begin{eqnarray}\label{15a}
f^{10}(x)&=&0.7539
\tanh\frac{x}{\sqrt{2}}+0.1005\tanh^{3}\frac{x}{\sqrt{2}}+0.0414\tanh^{5}\frac{x}{\sqrt{2}}+{}\nonumber\\
&&{}+0.0249\tanh^{7}\frac{x}{\sqrt{2}}+0.0178\tanh^{9}\frac{x}{\sqrt{2}}+0.0141\tanh^{11}\frac{x}{\sqrt{2}}+{}\nonumber\\
&&{}+0.0118\tanh^{13}\frac{x}{\sqrt{2}}+0.0103\tanh^{15}\frac{x}{\sqrt{2}}+0.0092\tanh^{17}\frac{x}{\sqrt{2}}+{}\nonumber\\
&&{}+0.0084\tanh^{19}\frac{x}{\sqrt{2}}+0.0078\tanh^{21}\frac{x}{\sqrt{2}}
\end{eqnarray}}
For $N\geq 10$, the relevant error is less than $0.5\%.$ When $ N$
increases, the relevant error will decrease. One can in principle
reach an arbitrary accuracy if one preserves sufficient terms in
equations (\ref{12a}) and (\ref{13a}) .

In the region $x\gg 1$, the solution (\ref{6a}) can be written as
\begin{equation}
 f(x)=1-b\e^{-\sqrt{2}x}+o(\e^{-\sqrt{2}x})
\end{equation}
where
\begin{equation}
b=\sum_{n=0}^{\infty}2(2n+1)c_{n} =\lim_{N\rightarrow\infty}
\sum_{n=0}^{N}2(2n+1)c_{n}^{(N)}.
\end{equation}
\section{The metric around a global monopole}
\noindent  The most general static metric with spherical symmetry
can be written as
 \begin{equation}\label{18a}
\d s^{2}=B(r)\d t^{2}-A(r)\d
 r^{2}-r^{2}(\d\theta^{2}+\sin\theta^{2}\d\varphi^{2}).
  \end{equation}

  The non-vanishing components of the Ricci tensor for this metric are
  \begin{eqnarray}
 &&R_{tt}=-\frac{B^{\prime\prime}}{2A}+\frac{B^{\prime}}{4A}\left(\frac{A^{\prime}}{A}+\frac{B^{\prime}}{B}\right)
 -\frac{1}{r}\frac{B^{\prime}}{A}\label{19a}\ ,\\
 &&R_{rr}=\frac{B^{\prime\prime}}{2B}-\frac{B^{\prime}}{4B}\left(\frac{A^{\prime}}{A}+\frac{B^{\prime}}{B}\right)
 -\frac{1}{r}\frac{A^{\prime}}{A}\ ,\\
 &&R_{\theta\theta}=-1+\frac{r}{2A}\left(-\frac{A^{\prime}}{A}+\frac{B^{\prime}}{B}\right)+\frac{1}{A}\ ,\\
 &&R_{\varphi\varphi}=\sin^{2}\theta R_{\theta\theta}.
 \end{eqnarray}
 The energy-momentum tensor of the monopole is given by
 \begin{eqnarray}
 &&T^{t}\,_{ t}=\eta^{2}\frac{f'^{2}}{2A}+\eta^{2}\frac{f^{2}}{r^{2}}+\frac{1}{4}\lambda\eta^{4}(f^{2}-1)^{2}\ ,\\
 &&T^{r}\,_{r}=-\eta^{2}\frac{f'^{2}}{2A}+\eta^{2}\frac{f^{2}}{r^{2}}+\frac{1}{4}\lambda\eta^{4}(f^{2}-1)^{2}\ ,\\
 &&T^{\theta}\,_{\theta}=T^{\varphi}\,_{\varphi}=\eta^{2}\frac{f'^{2}}{2A}+\frac{1}{4}\lambda\eta^{4}(f^{2}-1)^{2}.
\end{eqnarray}

  In the flat space the monopole core has size
 $\delta\sim(\eta\sqrt{\lambda})^{-1}$.
 For $\eta\ll m_{\mathrm{p}}$, where $ m_{\mathrm{p}}$ is the planck mass, we expect that
 gravity does not substantially change the structure of the monopole
 at small distance so that the flat space estimation of $\delta $
 still applies.

 The useful Einstein equations are given by
 \begin{eqnarray}\label{21a}
\frac{1}{A}(\frac{1}{r^{2}}-\frac{1}{r}\frac{A^{\prime}}{A})-\frac{1}{r^{2}}=8\pi
GT^{t}\,_{t}\ ,
\end{eqnarray}
\begin{eqnarray}\label{21b}
 \frac{1}{A}(\frac{1}{r^{2}}+\frac{1}{r}\frac{B^{\prime}}{B})-\frac{1}{r^{2}}=8\pi
 GT^{r}\,_{r}.
\end{eqnarray}
From equation (\ref{21a}), we get the general relation for $ A(r)$,
\begin{equation}\label{22a}
A^{-1}(r)=1-\frac{8\pi
G}{r}\int_{0}^{r}T^{t}\,_{t}r^{2}{\mathrm{d}r}
\end{equation}
and from equation (\ref{21b}), we get the general relation for $
B(r)$,
 \begin{equation}\label{23a}
B(r)=A^{-1}(r)\exp(-8\mu\int_{0}^{r}f^{2}r{\mathrm{d}r}).
\end{equation}
Substituting the solution $ f(r)$ into (\ref{22a}) and (\ref{23a}),
we get the solutions for $A(r)$ and $B(r)$. In the linear
approximation, i.e,  assuming that both $A(r)$ and $B(r)$ are very
close to unity, the result is
\begin{eqnarray}
&&A(r)=1+8\mu+\frac{Gm}{r}\ ,\\
&&B(r)=1-8(1+\beta)\mu-\frac{Gm}{r}.
\end{eqnarray}
where $\mu=\pi G\eta^{2}$,
 and
\begin{eqnarray}
&&m=8\pi\eta^{2}\int_{0}^{\infty}\left({f^{\prime}}^{2}+\frac{f^{2}-1}{r^{2}}+\frac{1}{4}\lambda\eta^{2}(f^{2}-1)^{2}\right)r^{2}{\mathrm{d}r}\\
&&\beta=\int_{0}^{\infty}{f^{\prime}}^{2} {\mathrm{d}r.}
\end{eqnarray}
By using equation (\ref{15a}), we obtain the numerical result for $
m,$
\begin{equation}
m=-46.99\eta^{2}\delta.
\end{equation}

In flat space, the core of a global monopole has size $\delta\sim
\lambda^{-\frac{1}{2}}\eta^{-1}.$ We expect that gravity does not
notably change the structure of the monopole at small distance, so
that,
\begin{equation}
|m|\sim 10\eta^{2}\delta.
\end{equation}
 There is a tiny repulsive gravitational potential due to the
mass term. A freely moving particle near the core experiences an
outward proper acceleration:
 \begin{equation}
\ddot{r}=-\frac{Gm}{r^{2}}=\frac{G|m|}{r^{2}}.
\end{equation}

Since $|m|\sim 10\eta^{2}{\delta}$, the mass term $\sim
10\eta^{2}\delta/r$.  When $r\gg \delta$, the mass term may be
negligible. Neglecting the mass term and rescaling the variables $r$
and $t,$ we can rewrite the metric as
\begin{equation}\label{31a}
{\d
s^{2}}={\mathrm{d}t^{2}}-{\mathrm{d}r^{2}}-(1-8\mu)r^{2}({\mathrm{d}\theta^{2}}+\sin\theta^{2}{\mathrm{d}\varphi^{2}}).
\end{equation}
The metric (\ref{31a}) describes a space with a solid deficit angel.

\section{The geodesics}

\noindent Let us now write down the equations for the geodesics in
the metric (\ref{18a}). From
\begin{equation}
\frac{\mathrm{d}^{2}x^{\mu}}{\mathrm{d}p^{2}}+\Gamma_{\nu\lambda}^{\mu}\frac{\mathrm{d}x^{\nu}}{\mathrm{d}p}\frac{\mathrm{d}x^{\lambda}}{\mathrm{d}p}=0
\end{equation}
we have
\begin{eqnarray}
&&A(r)\left(\frac{{\mathrm{d}r}}{{\mathrm{d}p}}\right)^{2}=\frac{1}{B(r)}-\frac{J^{2}}{r^{2}}-E\label{32a}\ ,\\
&&r^{2}\frac{\mathrm{d}\varphi}{\mathrm{d}p}=J \label{33a}\ ,\\
&&\frac{\mathrm{d}t}{\mathrm{d}p}=\frac{1}{B(r)}\label{34a}.
\end{eqnarray}

\noindent Here $J$ and $E$ are integral constants. $J$ represents
the angular momentum of the trajectory and $E^{\frac{1}{2}}$ is the
ratio between the proper time along the trajectory and the affine
parameter $p,$  i.e. $\mathrm{d}s^{2}=-E \mathrm{d}p^{2}$.

 From equations $(\ref{32a})-(\ref{34a}) $the shape of the path,
$r=r(\varphi)$, can be found. It is given by
\begin{equation}\label{35a}
\left|\frac{\mathrm{d}\varphi}{\mathrm{d}r}\right|\equiv\bigtriangleup(r)=\frac{1}{r^{2}}A^{1/2}
(r)\left[\frac{1}{JB(r)}-\frac{E}{J^{2}}-\frac{1}{r^{2}}\right]^{-1/2}
\end{equation}
And we can parametrize the trajectory in terms of the distance of
closest approach to the core, $r_{0}$, instead of the angular
momentum $J$. Form
$\frac{\mathrm{d}r}{\mathrm{d}\varphi}|_{r_{0}}=0,$ we have
\begin{equation}
J=r_{0}\left(1/B(r_{0})-E\right)^{1/2}.
\end{equation}
Then equation (\ref{35a}) reduce to
\begin{equation}
|\frac{\mathrm{d}\varphi}{\mathrm{d}r}|\equiv\bigtriangleup(r)=\frac{r_{0}}{r}A^{1/2}
(r)\left[\frac{B(r_{0})}{B(r)}\frac{1-EB(r)}{1-EB(r_{0})}-\left(\frac{r_{0}}{r}\right)^{2}\right]^{-1/{2}}
\end{equation}

Consider a trajectory starting from a distance $L\gg r_{0}$, and
with velocity $ V $ such that $V^{2}\gg \mu$. Most of the deflection
imprinted upon the trajectory by the gravitational filed will occur
in the region $r\approx r_{0}$,  where
\begin{equation}\label{38a}
\triangle(r)\approx
\frac{r_{0}}{r}\frac{1}{\left[1-({r_{0}/r})^{2}\right]^{1/2}}\left[1+4\mu+\frac{Gm}{2r}
-\frac{Gm}{4 V^{2}}\frac{1}{r_{0}}\frac{1}{1+{r_{0}/r}}\right]
\end{equation}
to the first order of $\mu$ and ${{\mu/V}^{2}}$. Also $E=1-V^{2}$ to
this order . The deflection angle is
\begin{equation}\label{39a}
\varepsilon=\pi-\Delta\varphi
 \end{equation}  where
\begin{equation}\label{40a}
 \Delta\varphi=2\int_{r_{0}}^{\infty} \triangle(r){\mathrm{d}r}.
 \end{equation}
Equations $(\ref{38a})- (\ref{40a})$ yield
 \begin{equation}\label{41a}
 \varepsilon=-4\pi\mu-(1-\frac{1}{2V^{2}})\frac{Gm}{r_{0}}
\end{equation}
In our order of approximation, $r_{0}$ coincides with the impact
parameter of the trajectory. The analogous result to equation
(\ref{41a}) for a gauge string is $\varepsilon=-4\pi\mu$ with $\mu$
the string energy per unit length. The monopole case differs from
that of a gauge string by the dependence of $\varepsilon$ upon both
the impact parameter and the velocity of the trajectory.

In the case of ultrarelativistic particles $(V\approx1)$, the
deflection angel is
\begin{equation}\label{42a}
\varepsilon=-4\pi\mu\left[1+\frac{Gm}{8\mu\delta}\frac{\delta}{r_{0}}\right]
\end{equation}
where ${Gm/8\mu\delta}={\mathrm{O}(1)}.$ We argue that light rays
are deflected by various angels, dependent upon the impact parameter
$r_{0}$. If $r_{0}$ is a galaxy scale and $\delta $ is of the order
of grand unification scale, the second term in equation (\ref{42a})
can be neglected. In this case, the deflected angle is independent
of the impact parameter. However, if the impact parameter is of the
scale of a mini soliton star, the second term in equation
(\ref{42a}) cannot be neglected. Such an object can be formed if a
global monopole is swallowed by a mini soliton star.

  There is a threshold velocity $V_{0}$ above which the monopole
  acts as a convergent lens and below which the monopole acts as
  a divergent lens. By setting $\varepsilon=0$ in equation (\ref{41a}) one
  obtains
\begin{equation}
V_{0}^{2}=\left[2(1+\frac{4\pi\mu\delta}{Gm}\frac{r_{0}}{\delta})\right]^{-1}.
\end{equation}
If $r_{0}$ and $\delta$ are of the scale of mini soliton star,
$V_{0}$ is about half the speed of light. When a particle moves at
speed $V_{0}$ there is no net deflection.

\section{discussion}
\noindent The crucial question now is what is the expected density
of global monopoles. Barriola and Vilenkin [17] proposed
 global monopole-antimonopole annihilation as a possible extremely
 efficient mechanism for reducing their number. The energy of the
 pair $(M\overline{M})$ is $E\sim \eta^{2}R$, where $R$ is the $M\overline{M}$
distance. The attractive force acting on $ M $and $\overline{M}$ is
$F$ and is independent upon the distance. The repulsive
gravitational force $F_{{\mathrm{rep}}}$ has maximum effect at
distance $R\sim \delta$, where it is
\begin{equation}
F_{{\mathrm{rep}}}\sim \frac{Gm^{2}}{\delta^{2}}\sim
10^{3}\mu\eta^{2}
\end{equation}
Because $\mu\sim10^{-6}$ at the typical grand unification scale
$10^{16}$ GeV, the repulsive force $F_{{\mathrm{rep}}}$ can be
neglected according to the large attractive force $F$. The large
attractive force between global monopole and  antimonopole  suggest
that $M\overline{M}$ annihilation is very effective and that the
monopole overproduction problem may not exist.

%\begin{acknowledgments}
%\end{acknowledgments}

%Just because of unusual number of tables stacked at end

\end{document}